\documentclass[preprint]{aastex}

\newcommand\kms{km s${}^{-1}$}
\newcommand\m{$\mu$}
\newcommand\ab{$\sim$}
\newcommand\p{$\pm$}
 
\begin{document}
\title{The Optical and Near-Infrared Morphologies
of Isolated Early Type Galaxies}
 
\author{James W. Colbert\altaffilmark{1},
 John S. Mulchaey\altaffilmark{2}, Ann I. Zabludoff\altaffilmark{3}}
\altaffiltext{1}{Department of Astronomy, UCLA, Los Angeles, CA 90095,
e-mail: colbert@astro.ucla.edu}
\altaffiltext{2}{Observatories of the Carnegie Institution of Washington, 
813 Santa Barbara St., Pasadena, CA 91101, e-mail: mulchaey@ociw.edu}
\altaffiltext{3}{Astronomy Department and Steward Observatory,
University of Arizona, Tucson, AZ 85721, e-mail: azabludoff@as.arizona.edu}
 
\setcounter{footnote}{0}
\begin{abstract}
In order to study early type galaxies in their simplest environments,
we have constructed a well-defined sample of 30 isolated 
galaxies. The sample contains all RC3 early-type galaxies with no
other cataloged galaxy with known redshift lying within a projected radius of
1h$_{\rm 100}^{-1}$ Mpc and \p 1000 \kms (where we use the recessional 
velocities in the RC3).  We have obtained optical and
near-infrared images of 23 of the galaxies and of 
a comparison sample of 13 early-type
galaxies in X-ray detected poor groups of galaxies. We have applied
the techniques of unsharp masking, galaxy model division, and color
maps to search for morphological features that might provide clues
to the evolution of these galaxies.  Evidence for dust features is
found in approximately 75\% of both the isolated and group galaxies (17 
of 22 and 9 of 12, respectively). However, shells or tidal
features are much more prevalent in our
isolated sample than in our group sample (9 of 22 $= 41\%$
versus 1 of 12 $= 8\%$, respectively). 
The isolation and colors of these shell galaxies make it unlikely that
tidal interactions or asymmetric star formation are the causes of such
features. One model that is not ruled out is that mergers produce the 
shells.
If shells and dust are both merger 
signatures, the absence of shells in group ellipticals
implies that shells: 1) form more easily, 2) are younger, and/or 3) are
longer-lived in isolated environments. 
\end{abstract}


\section{Introduction}

In the coarsest description of the Hubble sequence, early type
galaxies (ellipticals, S0's) are characterized by featureless light
profiles, whereas the disks of spiral galaxies have distinct
structures like dust lanes and arms.  Nevertheless, there is evidence
for structure in some early-types.  Shells (or ripples) are
detected in at least 10\% of the ellipticals in
the sample of Malin \& Carter (1983).  Large dust lanes or patches are
observed in many early-type galaxies 
(Ebneter et. al. 1988)
and are suggested in others by IRAS detections at 60 and 100 \m m 
(Thronson \& Bally 1987; Marston 1988; Thronson et al. 1989; 
Knapp et al.  1989; Bregman et al. 1998).
One problem in determining the
incidence and causes of such features is accounting first for the role
of environment.  Most early types lie in clusters and rich groups of
galaxies --- hot, dense regions that may affect both the formation and
evolution of features.  It is therefore essential to identify a
control sample in truly isolated environments, where the local galaxy
density is lower than in even poor groups of galaxies.

We have compiled a sample of 30 isolated early type galaxies in the
nearby universe.  If we define isolation as having no massive
companion within $1h^{-1}$ Mpc, then no close passage with another
galaxy at a relative velocity of \ab 300 \kms 
(approximately the mean pair-wise velocity of galaxies; Marzke 
et al. 1995) could have occurred in
the last $\sim 3h^{-1}$ Gyr (unless it resulted in a merger).  To
detect features such as dust and shells, it is necessary to apply a
range of image processing techniques to optical and near-infrared
images, identifying which methods work best for each type of
structure.  In particular, we would like to know 1) the fraction of
dusty {\it isolated} early types, 2) the fraction of {\it isolated}
early types with shells, and 3) how the fractions
of galaxies with dust and with
shells/tidal tails vary with galaxy environment.

The fraction of isolated early types that are dusty and/or disturbed
offer clues to the evolution of these galaxies.  For instance, Malin
\& Carter (1983) find that shells or ripples occur roughly five times
more frequently in environments outside of rich clusters.
Ebneter et al. (1988) argue a possible trend for early-type galaxies
outside of clusters to be dustier than their cluster counterparts. 
These
results suggest that, outside of clusters, such features are either
easier to form, younger, or longer-lived.
However, both the 
Malin \& Carter (1983) and Ebneter et al. (1988) samples 
contained few isolated galaxies. 
A large study of truly isolated galaxies is required to properly address the
issue of environmental differences in the structure of early types.

The frequency of dusty and morphologically disturbed 
isolated galaxies 
 can also illuminate the cause of structure in early types. 
The origins of such features are currently unknown.  Shells, for
example, could result from a merger of galaxies (Quinn 1984; Hernquist
\& Spergel 1992), weak
tidal interactions between galaxies (Thomson \& Wright 1990; Thomson 1991), or
asymmetric star formation (Lowenstein, Fabian, \& Nulsen 1987). 
If shells are found in isolated galaxies, the
tidal interactions model could likely be ruled out.
If the colors of isolated galaxies with shells are similar to those without
shells, then the asymmetric star formation model is less compelling.
With a well-defined sample of isolated early types, it is possible to test the 
validity of different shell formation models for the first time in a simple
environment.

In this paper, we analyze the sample of 30 isolated early type
galaxies and a comparison sample of 13 poor group ellipticals.  We define
these samples in \S 2.  We discuss the optical and near-infrared observations
and data reduction in \S 3 and \S 4.  We then search these images
for structure. To assist in the detection of features, we use several
different analysis techniques, including unsharp masking, model
division and color mapping. We describe this data analysis in section
\S 5. We give our results and discuss their relevance to the
possibility of mergers in section \S 6. Finally, we present our
conclusions in \S 7. All distances and luminosities in this paper are
calculated assuming a H$_{\rm o}$ $=$ 100 km s$^{-1}$ Mpc$^{-1}$ and 
q$_{\rm o}$ $=$ 0.5 cosmology.

\section{Sample Definition}

Using the new availability of large online databases, we have compiled
a unique and well-defined sample of 30 isolated early-type
galaxies from the RC3 catalog (de Vaucouleurs et. al. 1991). These
galaxies have no cataloged galaxies with known redshifts within a
projected radius of 1 Mpc h$_{\rm 100}^{-1}$ and \p 1000 \kms.  
We include all
revised morphological Hubble types, T $\le$ -3.  We further
restrict our sample to galaxies with z $<$ 0.033, within which the
isolation can be reliably tested. Due to the
incompleteness of the RC3 catalog, we also inspect each galaxy in sky
survey images and compare them with group and cluster catalogs drawn
from magnitude-limited redshift surveys (NASA Extra-galactic Database
(NED); CfA Redshift Survey, Huchra 1995). Galaxies that were in known
group or clusters as of October 1995 are excluded from our
sample. Table 1 lists our final isolated early-type galaxy sample. The
most common galaxy name is given in column 1, followed by the J2000 right
ascension and declination in columns 2 and 3, the recessional velocity
in column 4 and RC3 Hubble Type in column 5. The final column contains
comments including recently (i.e. since October 1995) cataloged
galaxies that are within our isolated criterion (i.e., 
within a radius of 1 Mpc
h$_{\rm 100}^{-1}$ and a velocity of \p 1000 \kms).
We note that several of the isolated galaxies
occur at very low Galactic latitude. The high level of extinction in these
fields likely plays an important role in these objects being defined
as \lq\lq isolated\rq\rq \ .
Galaxies at very low Galactic latitude are noted in Table 1 
in the Comments column.

We also examined a sample of 13 poor group ellipticals for
comparison. These groups are all X-ray detected with spectroscopically
confirmed memberships, so they are likely to be bound systems and not
chance superpositions (cf. Zabludoff \& Mulchaey 1998).  The poor
group elliptical sample is given in Table 2.  Figure 1 is a
histogram comparing the absolute B-band magnitudes of our isolated and poor
group samples. Based on a K-S test, the absolute magnitude distributions
for the two samples are
indistinguishable. 
However, on average, the group galaxies are closer than the isolated
galaxies. Therefore, because both samples were observed under similar
conditions (see \S3 and \S4), the linear resolution of our images 
is better on average for the group galaxies (Figure 2). In addition, the
signal-to-noise is somewhat better for our group sample. Even for the 
nearest galaxies in our sample, we image each field out to a radius
of at least $\sim$ 100 h$_{\rm 100}^{-1}$ kpc.
Because shell and tidal features
are expected on scales of tens of kiloparsec, field of view should not
limit our ability to detect structure in the nearer objects. Thus, given
the improvements in signal-to-noise and resolution, we are probably more
sensitive to detecting structure in our group sample than in our isolated 
sample, a factor that does not affect our conclusions (\S7).

\section{Optical Observations and Data Reduction}

Optical images were obtained at Las Campanas and Palomar
Observatories. The Las Campanas images were taken with a Johnson
B and a Kron-Cousins R filter at the
40 inch telescope in October 1995 and February 1996. The detector was
a Tektronix 2048x2048 CCD with fairly flat response between B and
R. The pixel scale was 0.696 arcsec pixel$^{-1}$, giving a total field
of view of 23.8$'$ x 23.8$'$.  All of the images taken at Las Campanas
were taken under photometric conditions with the exception of the
group elliptical NGC 533.  The measured seeing was in the range
of $\sim$ 0.9--1.9 arcsec, except for the galaxy NGC 3209, which 
was observed at very high airmass.  For each galaxy, two 150 second
integrations were obtained in each band.  A total of 19 isolated
galaxies and 11 group ellipticals were observed during the two Las
Campanas observing runs.

Three isolated ellipticals and one group elliptical were observed at
the Palomar Observatory 60 inch telescope in November 1995.  The
filters used were the Johnson B and I bands.
The conditions were non-photometric
during this observing run, so no attempt was made to measure
magnitudes for these galaxies. However, the seeing was very good ($\sim$
0.8--1.2 arcsec).

The optical data were reduced using standard techniques in IRAF. The
bias level was determined from the overscan region and then subtracted
from the images. Flat-fielding was done with twilight flats, typically
5-10 flats for each filter on each night. The flats were medianed
together to improve the signal-to-noise ratio and then
normalized. Each image was then divided by the normalized flat field
and sky subtracted. We determined the sky level by examining several
regions on the CCD free of emission from both stars and galaxies. The
two 150 second images of each galaxy and band were combined in such a
way as to eliminate cosmic ray strikes. The images taken at
Las Campanas were photometrically calibrated using standard star
fields in Graham (1982).
Total magnitudes were measured with the program SExtractor (Bertin 
\& Arnouts 1996) using a method similar to that proposed by Kron (1980). 
A comparison of our derived magnitudes with those in the literature
suggests a median external error estimate of 0.15 mag.
 Figure 3 shows the B-R colors we derive for
both samples using the Las Campanas observations. The range of colors
are comparable for the two samples, 
suggesting similar stellar populations. 

\section{Near-Infrared Observations and Data Reduction}

We obtained near-infrared images of 9 isolated galaxies and 2 
group galaxies on two
observing runs
at the 100 inch telescope at Las Campanas in September 1995 and April
1996.  The detector was a Rockwell NICMOS-3 HgCdTe 256x256 array. The
pixel scale was 0.35 arcsec pixel$^{-1}$, giving a total field of view
of approximately 1.5x1.5 arcminutes.  The filter used 
has a central wavelength around 2.15 microns and a FWHM
of 0.32 microns (this band is often referred to as K$_{short}$ (K$_{\rm S}$)). 
The conditions during both observing runs were photometric.
 Seeing varied between
0.6 and 1.2 arcsec, averaging around 0.8-0.9 arcsec.  For each galaxy,
we adopted an observing procedure identical to that described in
Mulchaey et al. (1997).  To minimize systematic errors we took the
galaxy images in a dithered pattern, with 6 twenty second images for
each position.  An equal amount of time was spent on dithered sky
exposures, typically 3--4 arcminutes from the galaxy. Generally, 13 on-source
positions were obtained resulting in a total integration time of
approximately 26 minutes. In addition, we also obtained a frame far
off to the side of any bright galaxies in order to better estimate the
sky level.

The data were reduced following the techniques described in Mulchaey
et al. (1997).  Flat field images were created for each night using
all the sky exposures obtained that night. A sky frame was formed for
each pair of object exposures by median filtering the sets of sky
exposures taken before and after each object pair.  We then applied
the flat field to the object frames and subtracted the appropriate sky
frame.  After mosaicing the six images at each dithered position, we
mosaiced the images from all thirteen dithered positions together.  In
most cases, stars were used to register the images. However, in a few
cases, the galaxy center was used. Because the galaxy emission clearly
extends beyond the edges of our final near-infrared images, we have not
tabulated total K$_{\rm S}$ magnitudes.

\section{Data Analysis}

The search for dust and tidal features in elliptical galaxies has been
greatly enhanced by the development of several data analysis
techniques.  Here, we use three of the most powerful techniques for
revealing structure in elliptical galaxies: unsharp masking, galaxy
model division and color mapping.  For images taken at Palomar, we
substitute the I band image for the R band throughout the following
discussion.

\subsection{Unsharp Masking}
	
Searching for shells, ripples, tidal tails
and other faint features in the outer
regions of ellipticals is severely hampered by the strong radial
intensity gradient of the underlying galaxy, which 
can wash out low surface brightness
features. One way of
removing this unwanted background light is unsharp masking. This
procedure involves taking the original image of the galaxy, 
smoothing it until all small scale features are erased,
and then subtracting this ``unsharp'' picture from the original galaxy
image. Only small scale features, such as thin shells or tails, then
remain. For smoothing we follow the method of McGaugh \& Bothun (1990)
and use a median filter. This method does not produce good results for
galaxy centers where the brightness gradient is strong, but is
excellent for finding details at the outer edges. After testing a
range of median filter sizes, we found that a filter of 10-20
arcseconds generally worked best at revealing features.
 For the velocity range of our galaxies
(5000-10,000 \kms ), this filter size 
corresponds to a physical size of \ab 6 h$_{\rm
100}^{-1}$ kpc. In all cases where a feature is uncovered using the 
unsharp masking technique, we are able to 
identify the feature in the 
original broad-band image by adjusting the stretch of the display.

\subsection{Galaxy Model Division}

While unsharp masking works well for the outer regions of a galaxy,
the inner region is also of interest. Again we have the problem of
a severe radial gradient in light intensity. To attack this problem we
choose the method of model division, used successfully by Ebneter,
Djorgovski \& Davis (1988), to identify dust features in galaxies.
Using the ELLIPSE program contained in
IRAF's STSDAS package, we fit elliptical isophotes to all galaxies in
all available filters. Using these models and the IRAF program
BMODEL we then make an artificial copy of the galaxy.  To avoid
contamination by stars or background galaxies, ELLIPSE is set to
reject 10-20\% of pixels farthest from the median surface brightness
value as it samples each isophote.  While ELLIPSE has
difficulty modeling the inner 2-3 pixels, it
provides a good model for the galaxy overall.  To search for large variations
from simple elliptical isophotes, we divide the original galaxy image
by the model image and look for residual features.  While Ebneter,
Djorgovski \& Davis (1988) go through a process of smoothing the outer
regions of their models while leaving the centers alone, we skip that
step, choosing instead to handle the analysis of the outer regions
with the unsharp masking technique.

Several of our ellipticals
show an odd \lq\lq quadrupole\rq\rq \
 feature near the galaxy center
in the model division image.  A similar feature was seen by Ebneter et
al.  (1988).
After reprocessing the images with a slightly different
choice for the galaxy center, Ebneter et al (1988)
found that several of their images no
longer showed a quadrupole, but instead revealed a disk. From this
result Ebneter et al. (1988) concluded that it is likely that all
galaxies with quadrupole features have disks, which make the total
isophotes non-elliptical and cause a numerical instability in the
isophote fitting algorithm. In cases where the 
quadrupoles could not be removed, Ebneter et al (1988) concluded that 
the galaxies contained unresolved disks.
However, Lauer (1985) shows that boxiness in the
stellar orbits can also cause a quadrupole (in his case, a significant
fourth Fourier harmonic of the pixel intensities).  Model division of
the boxy, poor group elliptical NGC 4261 (Lauer 1985), for example,
produces a quadrupole image with a phase that aligns roughly with the
diagonal axis of the galaxy's box in the unsharp mask image (see
Figure 7). We do not try to discriminate
between quadrupoles resulting from boxiness or diskiness in this
paper.

Following Ebneter et al. (1988), we reprocess our images by varying
the galaxy center used in our ellipse model.  This approach is
successful in several cases, revealing apparently flattened, dusty
structures.  In addition to NGC 4261, we are unable to remove
quadrupoles from the isolated elliptical NGC 3332 and the poor group
elliptical NGC 5129.

In the case of the morphologically disturbed galaxy, IC 2637, 
we cannot obtain
an adequate model of the galaxy isophotes using ELLIPSE. Thus, we do not
apply the model division technique to this galaxy.

\subsection{Color Mapping}

Features in elliptical galaxies can also be revealed by color maps.
This technique has been used by many authors, including Ebneter \&
Balick (1985) and Regan, Vogel \& Teuben (1995). For galaxies for which we
possess data in B, R, and K$_{\rm S}$, we make three color maps: B divided by
R, R divided by K$_{\rm S}$,
 and B divided by K$_{\rm S}$. In all images we observe a reddening 
color gradient toward the center, a phenomenon
observed before and thought to be
associated with an increasing metallicity toward
the center of elliptical galaxies (e.g. Davies, Sadler \& Peletier 1993;
Trager et al. 2000). 
Due to the intensity
gradient of the galaxy, proper alignment of the various images is very
important.  Slight misalignment leads to obvious dipoles in the final
color map.  We are careful to match the pixel scale and resolution
of the images in the three different bands. We observe structure in the color
maps of 26 of the 34 galaxies that we observed in multiple bands.

\section{Results and Discussion}

The identification of morphological structures is largely a subjective
process.  For our purposes here, we define structure as any feature
that deviates from simple elliptical isophotes.  Each image was
examined for features independently by J. Colbert and J. Mulchaey.
Only features identified by both of these authors are considered real.
There is good
agreement between them, both authors identifying 54 separate features
while disagreeing on only four. The results of each technique are
given in Table 3 for the isolated sample and in Table 4 for the group
galaxies. A more detailed description of the features found in each
case is given in Appendices A and B, and a few example images are given in 
Figures 4--7.

All three of the data analysis techniques that we employ reveal features
in our early-type galaxy sample.  In general, two types of
features are found: 1) shell or tidal features visible near the
outskirts of the stellar distribution and 2) dust features. The
difference between a \lq\lq shell\rq\rq \ and a tidal feature is
ambiguous in these images, and thus
we do not make a physical distinction between these two
types of features in this paper.

Dust lanes and patches are the most common type of feature we find. While both
the model division and color map techniques are successful at
uncovering dust, color maps provide the most sensitive probe of dust
features. Our color maps reveal dust in approximately 75\% of the
galaxies observed, with similar detection rates for the isolated and
group samples. 
 The model
division technique reveals dust features in $\sim$ 48\% (10 of
21) of the isolated galaxies and 42\% (5 of 12) of the group
galaxies. In addition, one isolated galaxy and 
 two group galaxies display 
quadrupole features that could be associated either with boxiness or a central disk.
Counting these galaxies among those with detected dust
brings the percentage of galaxies 
displaying dust in the model division maps to $\sim$ 52\% (11 of 21) 
for the isolated sample and  
$\sim$ 58\% (7 of 12) for the group sample. All the dust
features detected in the model division maps are also detected in the
color maps. This result suggests that although the model division
method is not the best way to find dust in early-type galaxies, the
technique is useful, especially if only one band is available.
It is also worth noting that the majority of dust patches found are
very close to the center of the galaxy (usually within the inner 
h$_{\rm 100}$$^{-1}$ kpc) .
However, because the signal to noise ratio tends to be highest near a galaxy's
center, our survey is most sensitive to dust there.

To calculate the probability that the dust fractions are the same in the
two samples, we use the cumulative binomial probability distribution to find 
if there is a "parent" dust fraction consistent with both the isolated and 
group galaxy dust fractions. For example, suppose that $k_1/n_1$ is the
observed fraction of isolated galaxies that have dust and $k_2/n_2 <
k_1/n_1$ is the observed fraction of group galaxies that have
dust.  For each parent probability $p = 0$ to 1, we calculate
the probability $P_1$ that at least $k_1$ of $n_1$ isolated galaxes have dust,
\begin{equation}
P_1 \equiv \sum_{j=k_1}^{n_1} (n_1)p^j(1-p)^{n_1-j} =
I_p(k_1,n_1-k_1+1),
\end{equation}
where $I_p$ is the incomplete beta function.For the same $p$,
 we calculate the probability $P_2$ that at most
$k_2$ of $n_2$ group galaxies are dusty, 
\begin{equation}
P_2 \equiv 1 - I_p(k_2+1,n_2-k_2).
\end{equation}
The product of the probability distributions
$P_1(p)P_2(p)$ is the joint probability distribution, whose maximum tells
us the likelihood that both $k_1/n_1$ and $k_2/n_2$
were drawn from the same parent fraction. The probability that the dust 
fraction of the isolated sample is the same as that of the group 
sample is 0.3317 (i.e., not statistically different).

All the shell features that we uncover are discovered by applying the
unsharp masking technique to our R-band images. While some of 
these features are also visible in the B-band images, none are
detected in the K$_{\rm S}$ images. Particularly noteworthy is the
galaxy NGC 7010, which is known to contain prominent shells (McGaugh
\& Bothun 1990), but shows no evidence for any structure in our
K$_{\rm S}$ image (the shells are easily visible in our R and B images).
The failure to detect shells in the K$_{\rm S}$ band may be an
indication that such features are below the surface brightness limit
of the images.  For this reason, we exclude from the tidal
feature statistics the two galaxies that are only observed in K$_{\rm
S}$.

We detect shell/tidal features in 9 of the 22 isolated
galaxies ($\sim$ 41\%). (The percentage is roughly
the same for galaxies classified as 
ellipticals and lenticulars.) 
However, only one of the 12
group early-type galaxies shows evidence for 
shells ($\sim$ 8\%). We should be more 
sensitive to the presence of 
shells in the group sample given that this sample is closer on average 
than the isolated sample (see \S 2). Using the binomial probability
distribution discussed above, the probability that the fraction of galaxies
with shell/tidal features is the same for both samples is only 0.016.
The tendency for
shells to be found preferentially around ellipticals in low density
environments was first suggested by Malin \& Carter (1983), who noted
that most shell galaxies are either apparently isolated or in very poor
groups. As the groups in our sample are all X-ray detected, they are
likely considerably richer than the typical group in Malin \& Carter
(1983). On the other hand, our isolation criteria ensure that the isolated 
sample galaxies lie in more rarefied environments than is typical
for the shell galaxies in Malin \& Carter (1983).

Our image analysis demonstrates that isolated elliptical galaxies are not 
featureless: $\sim$ 75\% of our isolated sample have dust features and 
41\% show evidence for shell/tidal features.
Because the factors that may affect the
evolution of early types in {\it isolated} environments are different
than those in clusters and poor groups, the isolated sample is a
useful control when compared to denser environments.  The fraction of
dusty galaxies is similar for the group and isolated samples.  The
shell discovery rate, on the other hand, is clearly dependent on 
environment, being significantly lower in the poor group
sample. 

For the isolated galaxies, tidal interactions with other 
galaxies (Thomson \& Wright 1990; Thomson 1991) are an unlikely
mechanism for shell formation as such features are not expected to last more
than $\sim$ 1 h$^{-1}_{\rm 100}$ Gyr (Quinn 1984)
 and our isolation criteria exclude 
encounters within the last $\sim$ 3 h$^{-1}_{\rm 100}$ Gyr.
The consistancy of the B-R colors of the isolated galaxies with and
without shells (B-R = 1.45  vs B-R= 1.44, galaxies with A$_{\rm B}$ $\le$ 0.3
 only) argues against the picture in which
shells are caused by asymmetric star formation (Lowenstein et al. 1987). 
However, galaxy-galaxy mergers 
are still a viable option.
If mergers are responsible for introducing dust into early-type
galaxies (i.e. Morganti et al. 1997), than the similar dust fractions
in our isolated and group samples suggest mergers have been important
in both environments.

If shells and tidal features are also an
indication of mergers,
then the differences between the isolated and group shell fractions
means
 we need to explain the differences between the
merger histories of galaxies in these two different environments. One
possibility is that dust features are produced by a wide range of
galaxy-galaxy mergers, while shells form in more limited types of
mergers that do not occur in denser environments. For example, shells
may form when a gas rich dwarf is swallowed by a larger galaxy (Weil
\& Hernquist 1993). However, gas rich dwarfs may not exist near the
centers of X-ray detected groups, where the intragroup medium and galaxy
number densities are highest. A second possibility is that 
even if shells do form in X-ray
group galaxies, the lifetime of such features is reduced by the hotter,
denser environment.  A third possibility is that because 
extended tidal/shell features are
expected to only have a limited lifetime (\ab 10$^9$ years, Quinn
1984), the difference in the isolated and group environments 
simply reflects that the isolated ellipticals have interacted more
recently. However, the mean colors (B-R) of the isolated and group
samples are similar (see Figure 3), so if interactions have occurred
more recently in the isolated ellipticals, there has been very little
if any effect on the stellar populations of the galaxies.

It is also interesting to note that all of the galaxies that display
shell/tidal features also contain dust. Given the dust and shell
detection rates for the isolated sample (77\% and 41\%, respectively),
we would expect $\sim$ 32\% (i.e., 0.77 $\times$ 0.41) or 7 of the 
22 isolated galaxies 
to display both dust and shells. The actual number of isolated galaxies with 
observable dust and shells is 9, which is not different from 7 at a 
statistically significant level. Thus, we are unable at the present time
to determine if dust and shells are correlated for isolated early-type
galaxies.  

\section{Conclusions}

We use multi-color images of a sample of isolated, early type galaxies
to study the properties of
such galaxies outside the hot, dense environments of rich
clusters. We compare this sample to a sample of early types in 
X-ray detected poor groups. Our survey has the important advantage over previous
studies that our strict isolation criteria ensures that no recent 
galaxy or environment interaction is possible except for a complete
merger. 

The majority of galaxies in our sample show evidence for
dust ($\sim$
75\%; 17 of 22 for the isolated sample and 
9 of 12 for the group sample).
We apply two different techniques to search for dust features,
model division and color maps. Both methods suggest that the incidence
of dust is comparable in isolated and poor group ellipticals.
However, the color maps are more successful at uncovering 
dust (approximately one-third of the galaxies that show evidence for dust
in the color maps show no evidence for dust in the model division images).
The dust detection rates are identical for our B-K$_{\rm S}$ and 
B-R images. Thus, although dust features tend to be most prominent 
in optical/near-infrared color maps, the optical
images alone appear adequate for identifying dust features. 

We have used the technique of unsharp masking to search for 
shells and tidal features.  We found these features in $\sim$ 41\%
(9 of 22) of
our isolated galaxies and approximately 8\% (1 of 12) of our 
group galaxies. The isolation and colors of the shell galaxies suggest
that neither tidal interactions nor asymmetric star formation cause such 
features. A more likely model is that mergers produce the shells.
Although the dust fraction is virtually identical for
the isolated and group environments,
the shell fractions are significantly different.
If shells and dust are
both merger signatures, the absence of shells in
 group ellipticals implies that
shells: 1) form more easily, 2) are younger, and/or 3) are longer-lived in
more isolated environments. 

We thank Chuck Keeton for providing useful information on statistics and 
the anonymous referee for comments that improved this paper. We would also like to thank Tod Lauer 
for his helpful suggestions.
This research involved the use of the NED database. 
Partial support for this program was provided from NASA grants
NAG 5-3529 and HF-01087.01-96A.

\vfill\eject

\appendix

\section{Comments On Individual Galaxies -- Isolated Galaxies}

NGC 179-- The B/R color map reveals large dust patches around the
center of the galaxy. IRAS
detected this galaxy at 60 and 100 $\mu$m (Knapp et al. 1989).

NGC 766--The B/R image of this galaxy reveals an unusually red core,
but does not appear to contain any obvious dust patches. It is also a
weak radio galaxy.

NGC 1132--No features are found in this galaxy. ASCA observations
indicate that this galaxy is surrounded by a large group-like 
X-ray emitting halo
(Mulchaey \& Zabludoff 1999).

A 0300+16--No features are found in this galaxy, which has previously
been identified as a FR I radio galaxy (Smith \& Heckman 1989).

UGC 02748--The B/R color map reveals several small dust patches,
including a strong one near the center. This galaxy is a known FR II
radio galaxy. 

A 0356+10--No features are detected in this FR II radio galaxy.

II Zw 017--The R/K$_{\rm S}$ color map reveals an asymmetric, non-elliptical
red feature running through the center, which is probably a dust
lane.

NGC 2110--The B/R image reveals dramatic dust lanes near the
galaxy's center. These dust lanes have previously been seen in HST
images (Mulchaey et al. 1994).  IRAS detected this Seyfert 2 galaxy
at 12, 25, 60 and 100 $\mu$m.

A 0718-34-- Unsharp masking reveals a shell/tidal feature to the southeast. The B/R
maps show several clear dust patches in the center. This
galaxy is a double radio source with several unresolved components (Jones \&
McAdam 1992).

NGC 3209-- Our images for this galaxy were taken under very poor seeing
conditions ($\sim$ 3$''$). Thus, although we 
found no evidence for features in this weak radio galaxy, our limits on 
structure are not very strong.

NGC 3332--This galaxy is clearly disturbed in our R image with a large
arm-like feature extending north of the galaxy. It is probably this
feature that has led to this galaxy's classification as an S0. While
we can not rule out the possibility that this feature is an arm, in
the unsharp masked image it looks very sharp and thus
more like a shell.
The R model divided image shows a quadrupole feature. The R/K$_{\rm
S}$ color map reveals red patches near the galaxy center which are
consistent with dust.
This galaxy was detected by IRAS at 60 and 100 $\mu$m
(Knapp et al. 1989).

IC 2637 (MRK 732)--IC 2637 is the most morphologically disturbed galaxy
we observed. In R band it looks more irregular than elliptical. In the
K$_{\rm S}$ band at low stretches, a disk or ring structure is
evident. At more saturated stretches, the infrared image does resemble
an elliptical.  All color maps show a strong dust lane running across
the galaxy. There are also several smaller red patches that may 
be associated with dust. This dust appears to be the major reason for
the galaxy's irregular appearance in the optical bands.  The unsharp
mask reveals structure on the north side of the galaxy that
may be shells or tidal tails. To the south is a large arc, far
enough away to avoid any confusion with the complicated structure of
the rest of the galaxy.
The morphology of IC 2637 is so distorted that we could not make an adequate
model of the galaxy isophotes using the task IRAF Ellipse. Thus, we did not
apply the model division technique for this object.
This Seyfert 1.5 galaxy (Hewitt \&
Burbidge 1991) was detected by IRAS at 12, 60 and 100 $\mu$m (Moshir et al. 
1990).

IC 2980-- Color maps reveal dust patches around the nucleus.

ESO 505- G 015-- Unsharp masking reveals shells to the southeast and
northwest. The B/K$_{\rm S}$ map indicates a possible dust patch
north of the nucleus.

ESO 065-01-- Unsharp masking reveals at least one shell off the major
axis of the elliptical. Unfortunately the other side of the elliptical
has a bad CCD column nearby, obscuring our ability to see structure on
that side. The R/K$_{\rm S}$ color map reveals two red patches
situated very assymetrically. The K$_{\rm S}$ model divided image does
not show any clear structure or bright spots associated with these
patches. Dust is therefore a good candidate. In the radio this is a
complex source with a strong core and edge darkened lobes(Jones \&
McAdam 1992).

ESO 574- G 017--This galaxy is a clearly disturbed, with faint
structure extending from it in a bow shock shape. The unsharp masking
reveals significant structure. On the east side the structure is close
to the galaxy and looks shell-like, whereas on the west side the
features are extending away from the galaxy like a tidal tail. The
color map reveals several red patches near the center of the galaxy,
which are likely due to dust. IRAS detected this galaxy at 25, 60 and 100
$\mu$m (Thronson et al. 1989).

IC 1156--This galaxy is only observed in the K$_{\rm S}$ band, so
we did not apply our three analysis techniques to this object.

NGC 6172--Unsharp masking reveals what appears to be a shell to the
southeast of NGC 6172, but not very far from its center ($\sim 25$
arcsec or 9 h$_{\rm 100}^{-1}$ kpc). 
The color maps reveal dust near the galaxy center. This galaxy was 
detected by IRAS at 25 and 60 $\mu$m (Moshir et al. 1990).

NGC 6799--The color maps reveal a large dust lane along the eastern
edge of the galaxy.
  
NGC 7010--This galaxy has prominent shells that have been reported previously
by McGaugh \& Bothun 1990. IRAS detected NGC 7010 at 12 and 100 $\mu$m
(Moshir et al. 1990).

IC 1392--Unsharp masking suggests a shell to the southwest. The B/I
image shows a small patch of dust at the galaxy's center.

IC 5258--There is evidence for dust near the center of the B/I image.

NGC 7618-- The B/I color image reveals a strong dark dust lane running
down the center.

\section{Comments on Individual Galaxies -- Poor Group Ellipticals}

NGC 383--A linear dust feature is seen in the color map. IRAS detected this
galaxy at 60 and 100 $\mu$m (Knapp et al. 1989).

NGC 533--We detect no features in this galaxy.

NGC 2563--Forbes \& Thomson (1992) have detected possible shells in
this galaxy, but no such feature is found in our somewhat shallower
image.  The color map reveals a large dust patch near the center of
the galaxy.

NGC 3091--This is the most luminous galaxy in HCG 42. No dust or tidal
features are detected.

NGC 3557--The color map reveals a possible ring of dust near the
center of the galaxy. NGC 3557 is a double tailed radio source with
a central knot and a jet (Birkinshaw \& Davies 1985). IRAS detected this
galaxy at 12, 60 and 100 $\mu$m (Knapp et al. 1989).

NGC 4261--The color map shows two large dust patches on either side of
what appears to be a bluer central lane.  A nuclear disk with a
similar orientation, but on a smaller scale (\ab 2$\arcsec$ or 200
h$_{\rm 100}$$^{-1}$ pc versus our \ab 8-10$\arcsec$ or 1 h$_{\rm
100}$$^{-1}$ kpc structure) has been reported by Jaffe et
al. (1996) based on HST images.  
Because the phase of the quadrupole in the model division image
aligns roughly with
the diagonal axis of the galaxy's box in the unsharp mask image,
the quadrupole 
is likely to result from the galaxy's boxiness (our Figure 7;
Lauer 1985).  This
galaxy was detected by IRAS at 12, 25, 60 and 100 $\mu$m (Knapp et
al. 1989).

NGC 4325--The color map reveals dust near the galaxy center.

NGC 4759--The second brightest galaxy in HCG 62. The color map reveals
strong dust patches near its center.

NGC 4761--The brightest galaxy in HCG 62. We find no evidence for dust
or tidal features, although the galaxy is part of an on-going
interaction with NGC 4759 (Hickson et al. 1992).

NGC 5044--The color map shows two large dust patches near the galaxy's
 center, with
a smaller patch off to the east. IRAS detected this galaxy at 12, 60 and
100 $\mu$m (Knapp et al. 1989).

NGC 5129--The model divided image of NGC 5129 has a strong quadrupole. The color map reveals
a bluer central lane surrounded by redder dust patches. A
possible tidal feature is detected near the southeastern corner of the galaxy.

IC 4296--The color map reveals large dust patches around the center
and some indication of a disk. IC 4296, a FR I radio galaxy, was
detected by IRAS at 60 and 100 $\mu$m (Knapp et al. 1989).

NGC 5846--While previous studies suggest NGC 5846 contains dust
(Goudfrooij \& Trinchieri 1998), our
analysis is incomplete because we only obtained a K$_{\rm S}$ image.

\vfill\eject

\newpage\begin{figure}
\epsscale{1}
\plotone{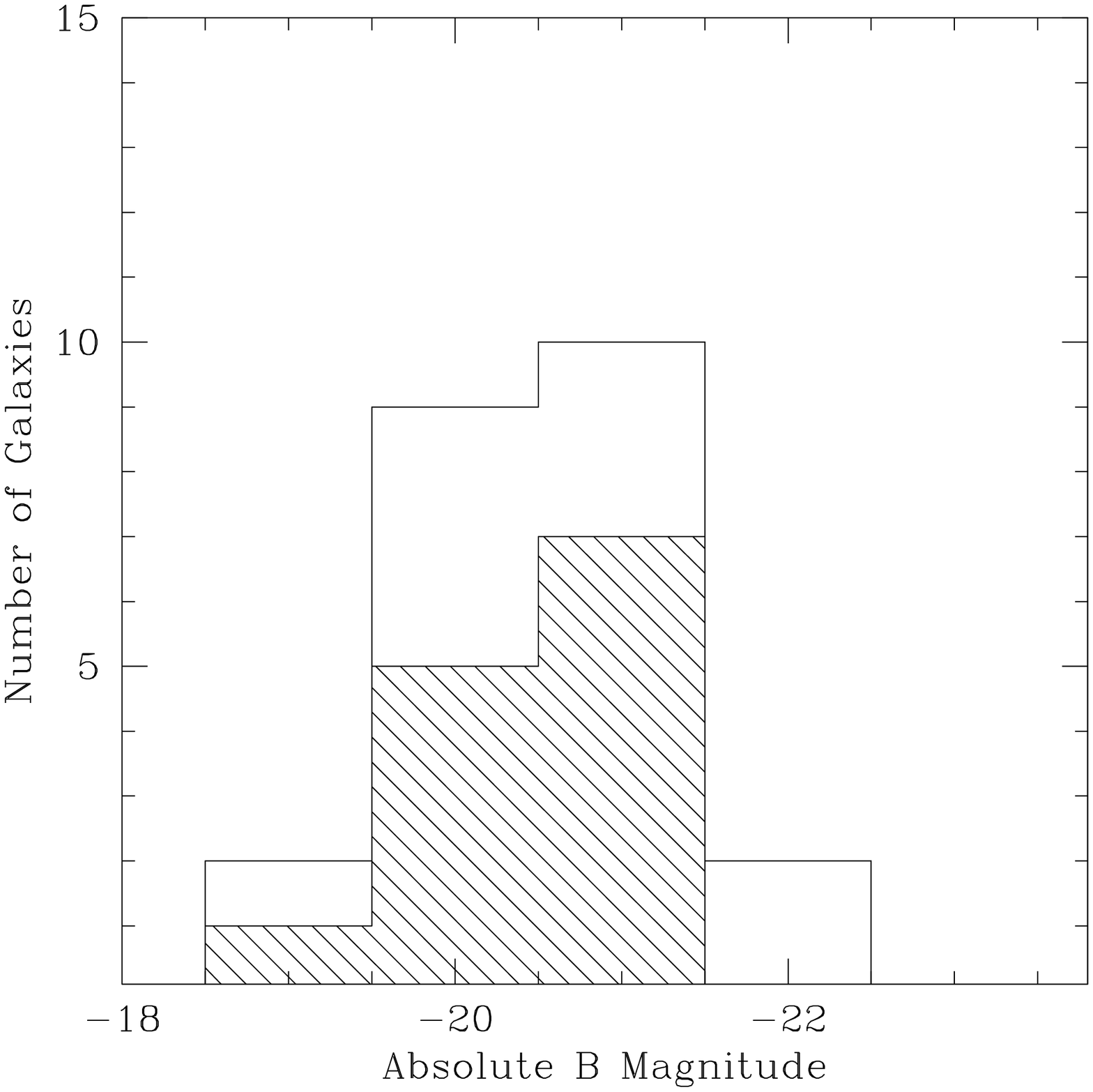}
\caption[JColbert.Figure1.ps]{Histogram of absolute B magnitude for the isolated early-type sample (open histogram) 
and the early-type galaxies in poor groups sample (dashed histogram).
A K-S test 
cannot distinguish between the two distributions. }
\end{figure}

\begin{figure}
\epsscale{1}
\plotone{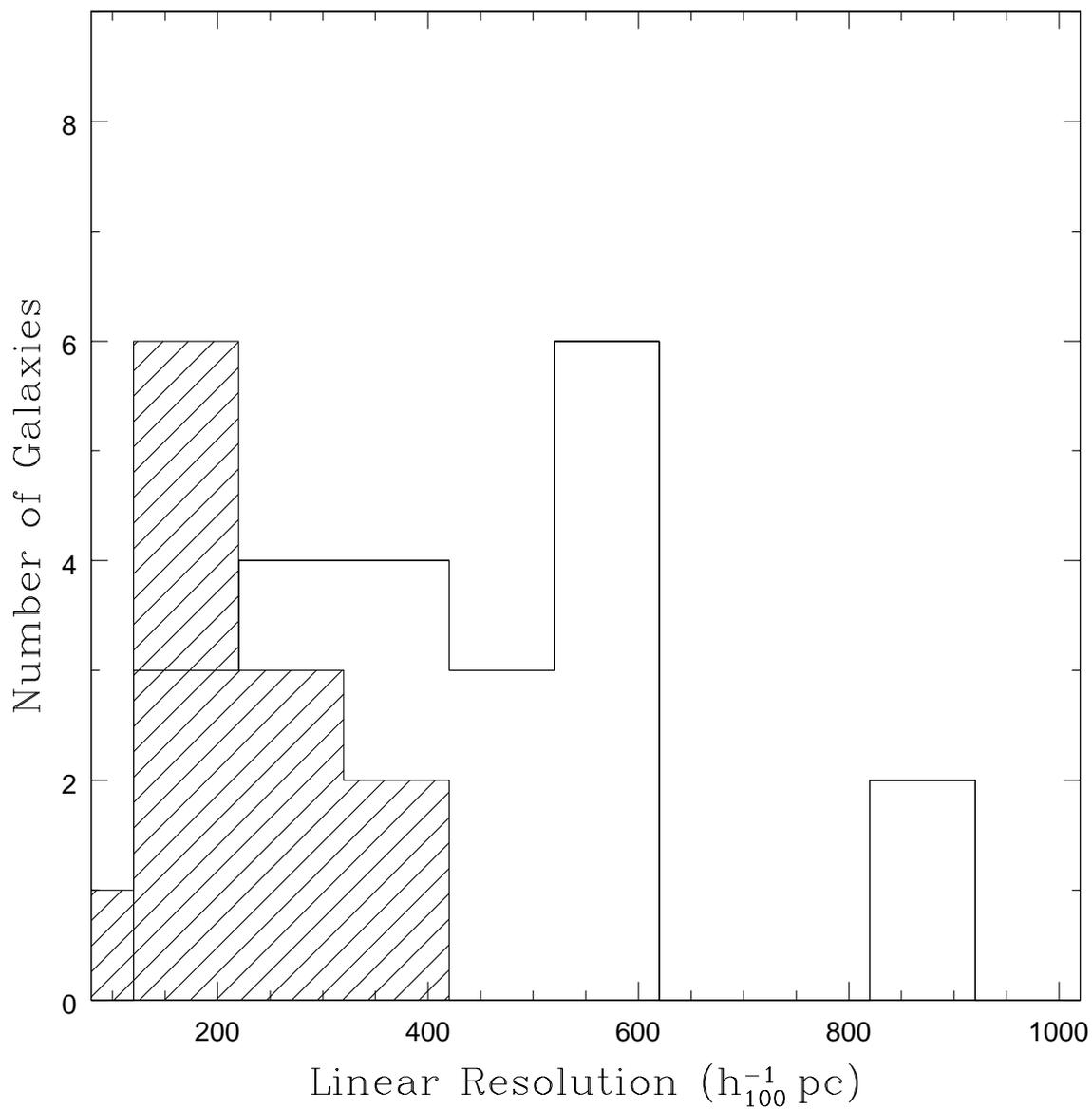}
\caption[JColbert.Figure2.ps]{Histogram of the linear resolution in
h$^{-1}$$_{\rm 100}$ parsecs
for the isolated early-type
sample (open histogram)
and the early-type galaxies in poor groups sample (dashed histogram).
The resolution is based on estimates of the seeing 
determined using the red image (either R
or I band). A K-S test indicates that the two distributions are different
at a significance of greater than 99\%.  
}
\end{figure}

\begin{figure}
\epsscale{1}
\plotone{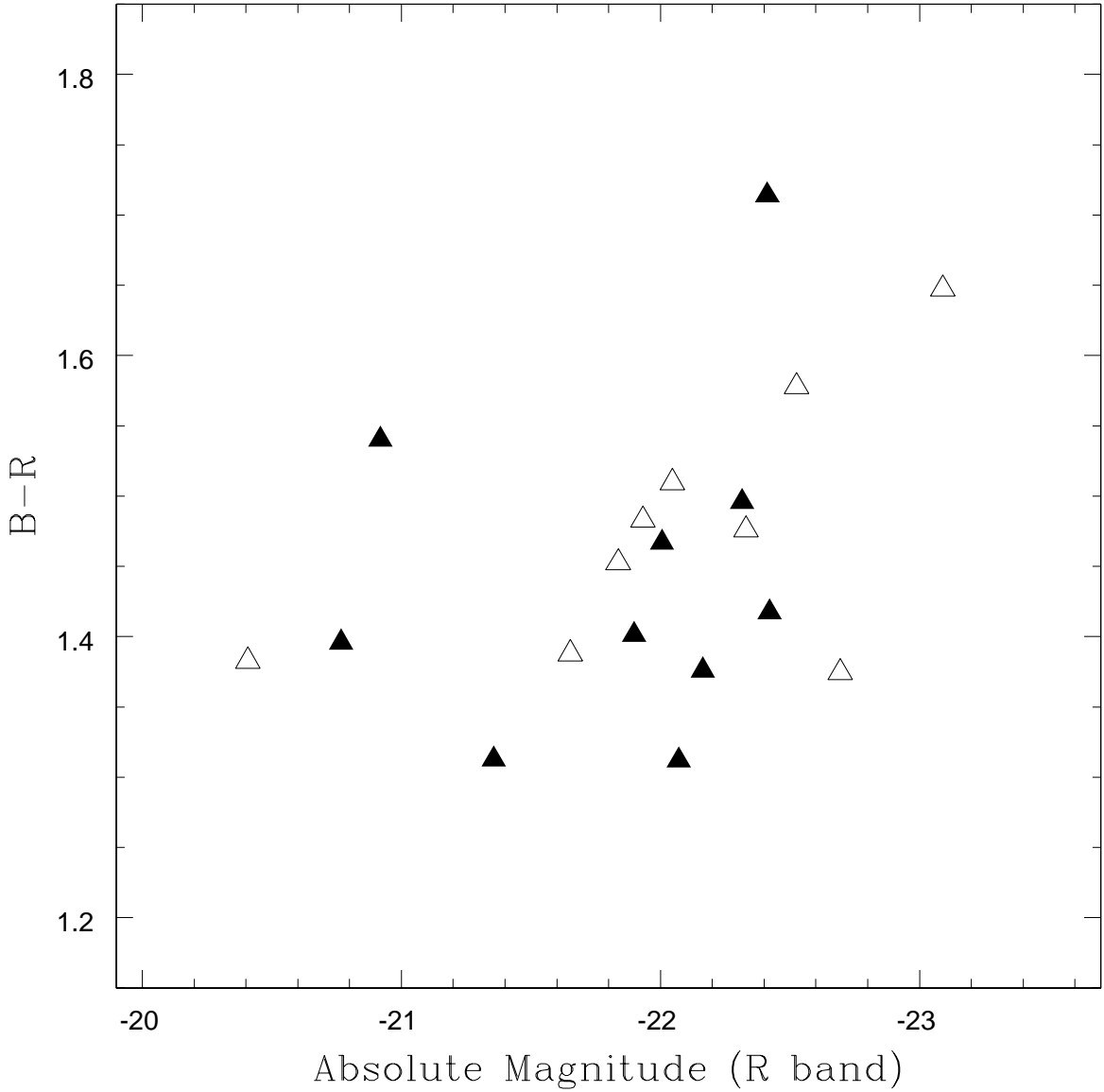}
\caption[JColbert.Figure3.ps]{Absolute R magnitude vs. B-R color diagram for the isolated 
early-type sample (open triangles) and the group early-type sample (filled
triangles). Only galaxies with Galactic extinction values less than
A$_{\rm B}$=0.3 are included.
The colors of the galaxies in the two
different environments are comparable, suggesting similar stellar 
populations in these samples.
 }
\end{figure} 

\begin{figure}
\epsscale{0.9}
\plotone{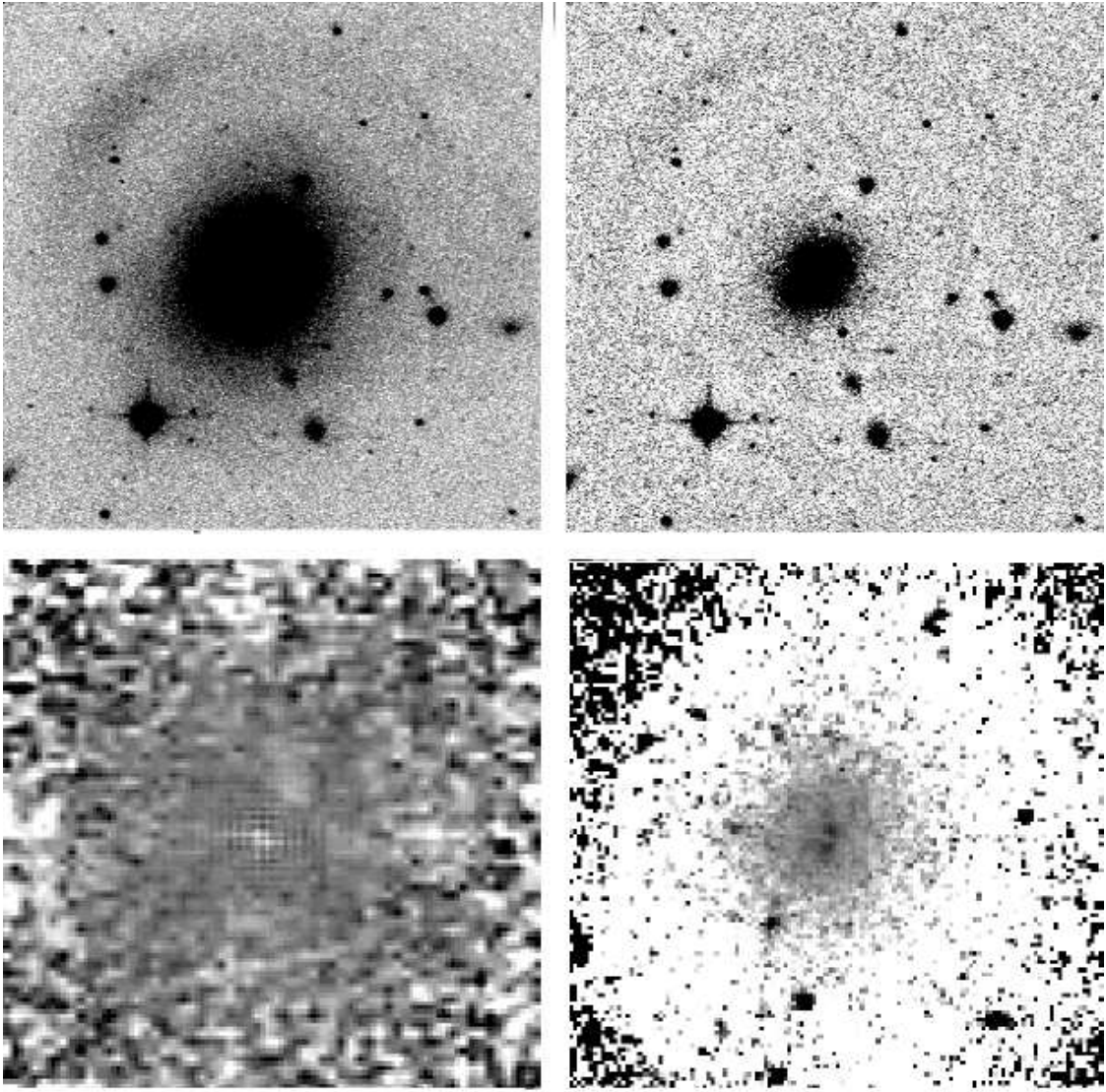}
\caption[JColbert.Figure4.ps]{The analyzed images for NGC 3332. The plotted images 
correspond to:
R-band (upper left), unsharp mask (upper right), R-band model division 
(lower left) and R/K$_{\rm S}$ (lower right), with scales on the sky of 
3.5$^{\prime}$ $\times$ 3.5$^{\prime}$, 3.5$^{\prime}$ $\times$ 3.5$^{\prime}$,  90$\arcsec$ $\times$ 90$\arcsec$, and 90$\arcsec$ $\times$ 90$\arcsec$ respectively. Notice the shell/tidal
features visible to the northeast of the galaxy in the unsharp 
masked image. The R-model image has a very weak 
quadrupole feature.  Large symmetrical dust features are present in the color map image. For the model division and color maps, darker pixels indicate redder regions.
}
\end{figure}

\begin{figure}
\epsscale{0.9}
\plotone{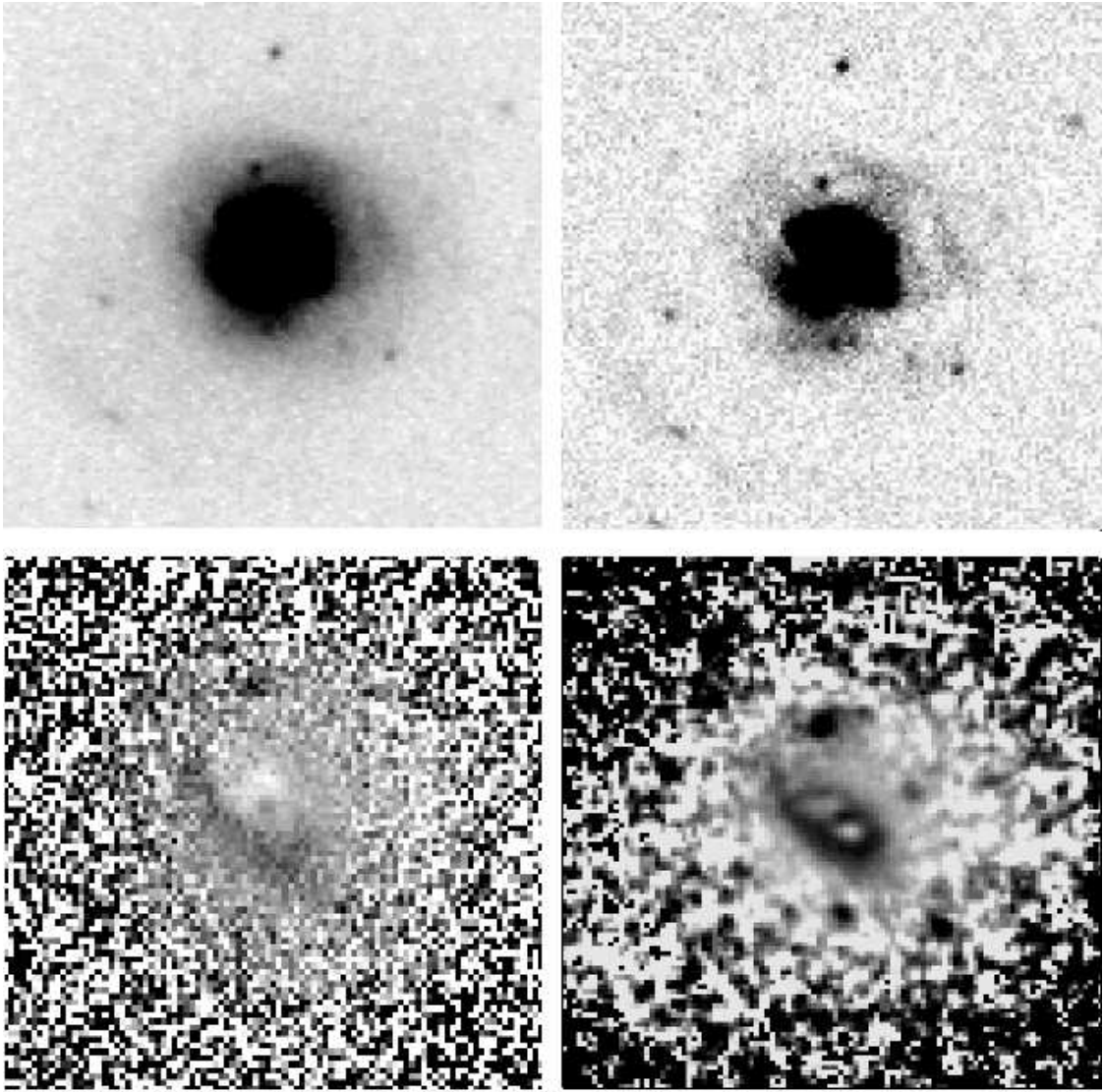}
\caption[JColbert.Figure5.ps]{The analyzed images for IC 2637. The plotted images 
correspond to:
R-band (upper left), unsharp mask (upper right), B/R
(lower left) and R/K$_{\rm S}$ (lower right), with scales on the sky of 
90$\arcsec$ $\times$ 90$\arcsec$, 90$\arcsec$ $\times$ 90$\arcsec$,  60$\arcsec$ $\times$ 60$\arcsec$, and 60$\arcsec$ $\times$ 60$\arcsec$ respectively. Note the prominent tidal
features and extensive dust structures. For the color maps, darker pixels indicate redder regions.}
\end{figure}

\begin{figure}
\epsscale{0.9}
\plotone{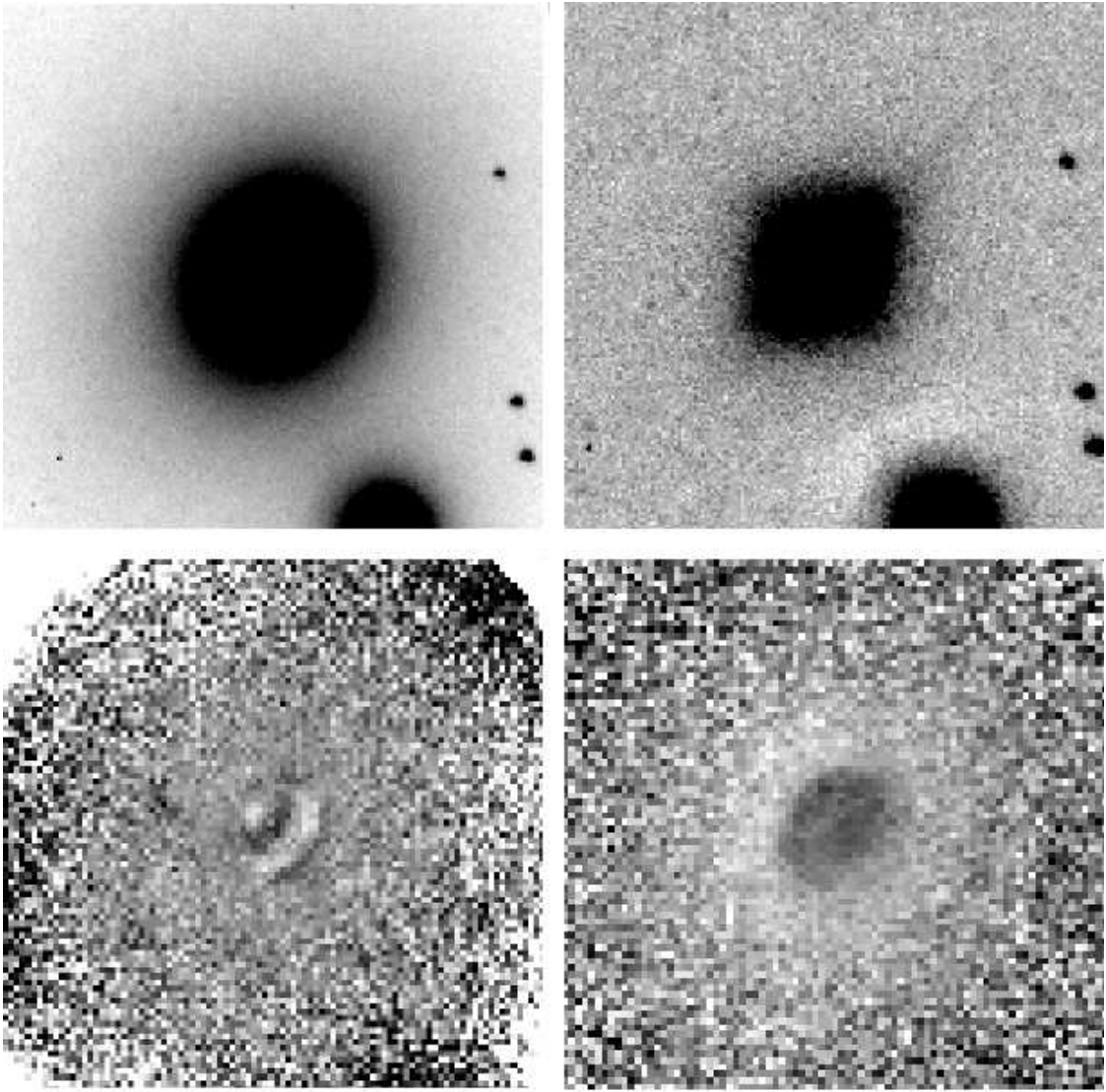}
\caption[JColbert.Figure6.ps]{The analyzed images for NGC 383. The plotted images
correspond to:
I-band (upper left), unsharp mask (upper right), I-band model division 
(lower left) and B/I (lower right), with scales on the sky of 
120$\arcsec$ $\times$ 120$\arcsec$, 120$\arcsec$ $\times$ 120$\arcsec$,  60$\arcsec$ $\times$ 60$\arcsec$, and 60$\arcsec$ $\times$ 60$\arcsec$ respectively. While the unsharp mask reveals 
no features, prominent dust structure is visible in both the model division
image and color map. For the model division and color maps, darker pixels indicate redder regions.}
\end{figure}

\begin{figure}
\epsscale{0.8}
\plotone{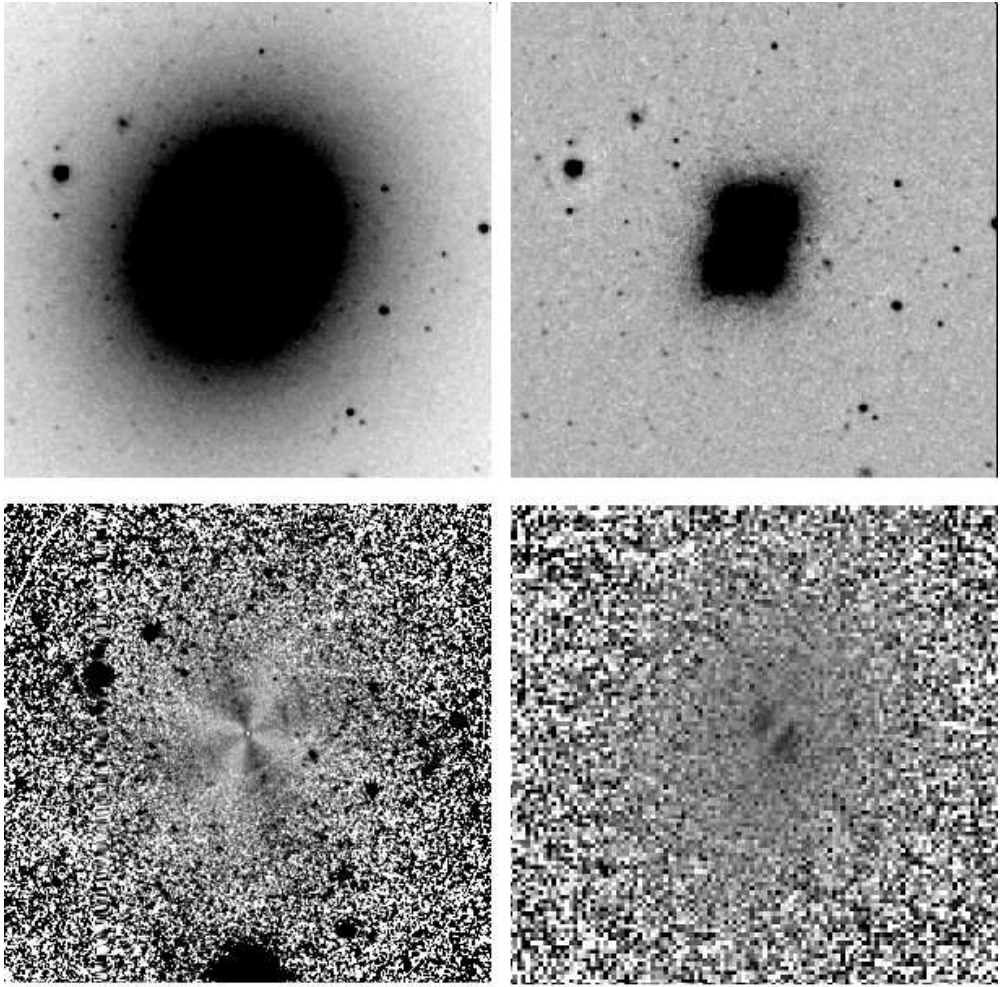}
\caption[JColbert.Figure7.ps]{The analyzed images for NGC 4261. The plotted images
correspond to:
R-band (upper left), unsharp mask (upper right), R-band model division 
(lower left) and B/R (lower right), with scales on the sky of 
3$^{\prime}$ $\times$ 3$^{\prime}$, 3$^{\prime}$ $\times$ 3$^{\prime}$,  3.5$^{\prime}$ $\times$ 3.5$^{\prime}$, and 90$\arcsec$ $\times$ 90$\arcsec$ respectively.
No features are seen in the unsharp mask image. NGC 4261 is clearly boxy, which is the
likely explanation for the quadrupole observed in the model-division
image (Lauer 1985).  Note how the phase of the quadrupole aligns
roughly with the diagonal axis of the galaxy's box in the unsharp mask
image.  With higher resolution images, several authors have detected a
central dust disk in this galaxy (Kormendy \& Stauffer 1986; Jaffe
et al. 1996).  The B/R color map also reveals complex structure
in the core.  For the model division and color maps, darker pixels indicate redder regions.}
\end{figure}

\begin{deluxetable}{rrrrrr}
\tablecaption{Isolated Early-Type Sample}
\tablehead{\colhead{Galaxy} & \colhead{$\alpha$(2000)} & \colhead{$\delta$(2000)} & \colhead{cz} & \colhead{Type} & \colhead{Comments}}
\startdata
NGC 179 & 00 37 46.2 & -17 50 57 & 6027& -3.0 & \nodata \\
NGC 766 & 01 58 42.0 & 08 20 48 & 8104 & -5.0 & \nodata \\
NGC 1132 & 02 52 51.8 & -01 16 28 & 6953 & -4.5 & NGC 1126 8.4$^{\prime}$ away; \\
& & & &  & group-like X-ray halo\tablenotemark{a} \\
A0300+16 & 03 03 15.0 & 16 26 19 & 9740 & -5.0 & \nodata \\
UGC 2748 & 03 27 54.2 & 02 33 42 & 9060 & -4.7 & poor group \\
A0356+10 & 03 58 54.4 & 10 26 02 & 9130 & -5.0 & low Galactic latitude\\
IIZw017 & 04 35 13.8 & -01 43 53   & 9730 & -5.0 & MCG +00-12-049 21.8$^{\prime}$ away\\
NGC 2110 & 05 52 11.4 & -07 27 22 & 2284 & -3.0 & Seyfert 2 galaxy; \\
& & & & & low Galactic latitude \\
A0718-34 & 07 20 47.5 & -34 07 07 & 8900 & -3.0 & low Galactic latitude\\
NGC 3209 & 10 20 38.5 & 25 30 15  & 6107 & -5.0 & poor group\tablenotemark{b}  \\
NGC 3332 & 10 40 28.6 & 09 10 59 & 5727 & -3.0 & \nodata \\
IC 2637 & 11 13 49.8 & 09 35 11 & 8763 & -4.0 & Seyfert 1.5 galaxy; \\
& & & & & FGC 148A 7.0$^{\prime}$ away\\
IC 2980 & 11 57 30.0 & -73 41 06 & 8352 & -5.0 & low Galactic latitude\\
ESO 505-G015 & 12 07 07.9 & -25 41 34 & 7467 & -4.0 & ESO 505-G017 
11.8$^{\prime}$ away \\
ESO 065-G001 & 12 37 12.0 & -72 35 31 & 7058 & -5.0 & WKK 1219 11.0$^{\prime}$
 away; \\
& & & & & low Galactic latitude\\
ESO 574-G017 & 12 40 35.0 & -20 33 45 & 8583 & -4.0 & \nodata \\
NGC 5413 & 13 57 53.5 & 64 54 40 & 9634 & -5.0 &  \nodata \\
UGC 9874 & 15 27 14.9 & 77 09 25 & 5380 & -5.0 & \nodata \\ 
UGC 10115 & 15 57 07.8 & 63 55 03 & 9216 & -5.0 & MCG +11-19-030 4.6$^{\prime}$ 
away \\
IC 1156 & 16 00 37.3 & 19 43 24 & 9475 & -5.0 & \nodata \\
NGC 6172 & 16 22 10.3 & -01 30 54 & 4979 & -4.0 &  \nodata \\ 
A1836+17 & 18 38 25.4 & 17 11 51 & 5090 & -3.0 & UGC 11301 21.4$^{\prime}$ away\\
NGC 6702 & 18 46 57.6 & 45 42 20 & 4727 & -5.0 & \nodata \\
NGC 6799 & 19 32 16.1 & -55 54 29 & 5590 & -3.7&  \nodata  \\
NGC 6849 & 20 06 16.2 & -40 11 51 & 6040 & -3.0 & \nodata \\
NGC 6944 & 20 38 23.8 & 06 59 47 & 4417 & -3.0 & NGC 6944A 6.3$^{\prime}$ away \\
NGC 7010 & 21 04 39.5 & -12 20 18 & 8486 & -4.0 & \nodata \\
IC 1392  & 21 35 32.6 & 35 23 55 & 4395 & -3.0 & UGC 11775 4.2$^{\prime}$ away; \\
& & & & & UGC 11781 
22.4$^{\prime}$ away; \\
& & & & & low Galactic latitude \\
IC 5258 &  22 51 31.6 & 23 04 50 & 7747 & -3.0 & \nodata \\
NGC 7618 & 23 19 47.3 & 42 51 08 & 5189 & -5.0 & UGC 12491 14.1$^{\prime}$ away; \\
& & & & & low Galactic latitude\\

\enddata
\tablerefs{(a) Mulchaey \& Zabludoff (1999); (b) Ramella et al. (1997)}
\end{deluxetable}

\begin{deluxetable}{rrrrrr}
\tablecaption{Group Early-Type Sample}
\tablehead{\colhead{Galaxy} & \colhead{$\alpha$(2000)} & \colhead{$\delta$(2000)} & \colhead{cz} & \colhead{Type} & \colhead{Comments}}
\startdata

NGC 383 & 01 07 24.9 & 32 24 45 & 5090 & -3.0 & \nodata \\
NGC 533 & 01 25 31.3 & 01 45 33 & 5544 & -5.0 & \nodata \\
NGC 2563& 08 20 35.7 & 21 04 04 & 4480 & -2.0 & \nodata \\
NGC 3091& 10 00 14.3 & -19 38 13& 3964 & -5.0 & \nodata \\
NGC 3557& 11 09 57.4 & -37 32 17& 3067 & -5.0 & \nodata \\
NGC 4261& 12 19 23.2 &  05 49 31& 2210 & -5.0 & \nodata \\
NGC 4325& 12 33 06.7 &  10 37 16& 7709 & -5.0 & \nodata \\
NGC 4759& 12 53 04.6 & -09 12 02& 3561 & -2.0 & member of HCG 62; \\
& & & & & paired with NGC 4761 \\
NGC 4761& 12 53 05.8 & -09 12 16& 4259 & -1.0 & member of HCG 62; \\
& & & & &  paired with NGC 4759 \\
NGC 5044& 13 15 24.0 & -16 23 06& 2704 & -5.0 & \nodata \\
NGC 5129& 13 24 10.0 &  13 58 36& 6908 & -5.0 & \nodata \\
IC 4296 & 13 36 39.4 & -33 58 00& 3761 & -5.0 & \nodata \\
NGC 5846& 15 06 29.2 &  01 36 21& 1822 & -5.0 & \nodata \\ 
\enddata
\end{deluxetable} 

\begin{deluxetable}{rrrrrrrr}
\tablecaption{Properties of Isolated Early-type Galaxies}
\tablehead{\colhead{Galaxy} & \colhead{Telescope\tablenotemark{a}} & \colhead{Bands} & \colhead{R} & \colhead{B} & \colhead{Unsharp} & \colhead{Model} & \colhead{Color} \\
& & & \colhead{Mag} & \colhead{Mag} & \colhead{Masking} & \colhead{Division} & \colhead{Map}}
\startdata
NGC 179   & L40\phs & R,B\phs & 13.20 & 14.63 &   no & no  & dust \\
NGC 766   & L40\phs & R,B\phs & 12.58 & 14.07 &   no & no  & no \\
NGC 1132  & L40\phs & R,B\phs & 11.97 & 13.49 &   no & no  & no \\
A0300+16  & L40\phs & R,B\phs & 14.46 & 15.96 &   no & no  & no \\
UGC 2748  & L40\phs & R,B\phs & 13.53 & 15.19 &   no & no  & dust \\
A0356+10  & L40\phs & R,B\phs & 14.20 & 15.77 &   no & no  & no \\
IIZw017   & L40\phs & R,B\phs & 14.17 & 15.79 &   no & dust & dust \\
NGC 2110  & L40\phs & R,B\phs & 11.36 & 13.39 &   no & dust & dust \\
A0718-34  & L40\phs & R,B\phs & 13.32 & 15.35 &   shells & dust & dust \\
NGC 3209  & L40\phs & R,B\phs & 11.68 & 13.21 &   no & no & no \\
NGC 3332  & L40,L100 & K$_{\rm S}$,R,B
 & 11.88 & 13.40 & shells & Quadrupole & dust \\
IC 2637   & L40,L100 & K$_{\rm S}$,R,B & 
12.72 &14.07 & shells & \nodata & dust \\
IC 2980   & L40,L100 & K$_{\rm S}$,R,B &
 11.58 &13.47 & no & dust & dust \\
ESO 505-G015 & L40\phs & R,B\phs & 12.51 &14.26 & shells  & no & dust \\
ESO 065-G001 & L40,L100 & K$_{\rm S}$,R,B &
 12.80 & 14.89 & shells & dust & dust \\
ESO 574-G017 & L40,L100 & K$_{\rm S}$,R,B & 
13.47 & 14.87 & shells & dust & dust \\
IC 1156   &  L100\phn & K$_{\rm S}$\phs & \nodata & \nodata &  \nodata & \nodata & \nodata \\
NGC 6172  & L40,L100 & K$_{\rm S}$,R,B &
 12.61 &13.98 & shells & dust & dust \\
NGC 6799  & L40,L100 & K$_{\rm S}$,R,B &
 12.01 &13.51 & no & dust & dust \\
NGC 7010  & L40,L100 & K$_{\rm S}$,R,B &
 12.37 &14.16 & shells & dust & dust \\
IC 1392   & P60\phs      &I,B\phs  & \nodata & \nodata &  shells & no & dust \\
IC 5258   & P60\phs      &I,B\phs  & \nodata & \nodata &  no & no & dust \\
NGC 7618  & P60\phs      &I,B\phs  & \nodata & \nodata &  no & dust & dust \\ 
\enddata

\tablenotetext{a} {L40= Las Campanas 40-inch; L100= Las Campanas 100-inch; P60= Palomar
60-inch.}
\end{deluxetable}

\begin{deluxetable}{rrrrrrrr}
\tablecaption{Properties of Early-type Group Galaxies}
\tablehead{\colhead{Galaxy} & \colhead{Telescope\tablenotemark{a}} & \colhead{Bands} & \colhead{R} & \colhead{B} & \colhead{Unsharp} & \colhead{Model} & \colhead{Color} \\
& & & \colhead{Mag} & \colhead{Mag} & \colhead{Masking} & \colhead{Division} & \colhead{Map}}
\startdata
NGC 383& P60\phs & I,B\phs & \nodata & \nodata & no & dust & dust \\
NGC 533& L40\phs &R,B\phs  & b\phs & b\phs & no & no & no \\
NGC 2563& L40\phs &R,B\phs & 11.73 &13.19 & no & no & dust \\
NGC 3091& L40\phs &R,B\phs & 10.59 &12.24 & no & no & no \\
NGC 3557& L40\phs &R,B\phs & 9.89  &11.43 & no & dust & dust \\
NGC 4261& L40\phs &R,B\phs & 9.73  &11.27 & no & Quadrupole & dust \\
NGC 4325& L40,L100& K$_{\rm S}$,R,B &
 12.61 &14.15 & no & no & dust \\
NGC 4759& L40\phs &R,B\phs & 12.50 &13.97 & no & dust & dust \\
NGC 4761& L40\phs &R,B\phs & 11.46 &13.00 & no & no & no \\
NGC 5044& L40\phs &R,B\phs & 10.02 &11.61 & no & dust & dust \\
NGC 5129& L40,L100 & K$_{\rm S}$,R,B &
 11.59 &13.01 & shells & Quadrupole & dust \\
IC 4296& L40\phs & R,B\phs &  9.96 & 11.71 & no & dust & dust \\
NGC 5846& L100\phn & K$_{\rm S}$\phs & \nodata & \nodata & \nodata & \nodata & \nodata \\
\enddata

\tablenotetext{a} {L40= Las Campanas 40-inch; L100= Las Campanas 100-inch; P60= Palomar
60-inch.}
\tablenotetext{b} {Data taken under non-photometric conditions, so no magnitudes are given.}
\end{deluxetable}

\end{document}